\documentclass{article}
\usepackage{spconf,amsmath,graphicx}
\usepackage{booktabs}
\usepackage{subcaption}
\usepackage[table]{xcolor}
\usepackage{comment}
\usepackage{standalone}
\usepackage{pgfplots}
\usepackage{tikz}
\usepackage{url}
\usepackage{hyperref}

\usepgfplotslibrary{fillbetween}

\title{Stable Distillation: Regularizing Continued Pre-training for Low-Resource Automatic Speech Recognition}
\name{Ashish Seth$^{1*}$ \quad Sreyan Ghosh$^{2*}$ \quad S. Umesh$^1$ \quad Dinesh Manocha$^2$ \thanks{$^*$ These authors contributed equally to this work.}}
\address{$^1$IIT Madras, India, $^2$University of Maryland, College Park, USA}
\begin{document}
%
\maketitle
\begin{abstract}
Continued self-supervised (SSL) pre-training for adapting existing SSL models to the target domain has shown to be extremely effective for low-resource Automatic Speech Recognition (ASR). This paper proposes Stable Distillation, a simple and novel approach for SSL-based continued pre-training that boosts ASR performance in the target domain where both labeled and unlabeled data are limited. Stable Distillation employs self-distillation as regularization for continued pre-training, alleviating the over-fitting issue, a common problem continued pre-training faces when the source and target domains differ. Specifically, first, we perform \textit{vanilla} continued pre-training on an initial SSL pre-trained model on the target domain ASR dataset and call it the teacher. Next, we take the same initial pre-trained model as a student to perform continued pre-training while enforcing its hidden representations to be close to that of the teacher (via MSE loss). This student is then used for downstream ASR fine-tuning on the target dataset. In practice, Stable Distillation outperforms all our baselines by 0.8 - 7 WER when evaluated in various experimental settings\footnote{\url{https://github.com/cs20s030/stable_distillation}}.

\end{abstract}

%
\begin{keywords}
speech recognition, self-supervised learning, self-distillation
\end{keywords}
\section{Introduction}
\label{sec:intro}

\begin{figure}[t]
    \centering
    \includegraphics[width=0.4\textwidth]{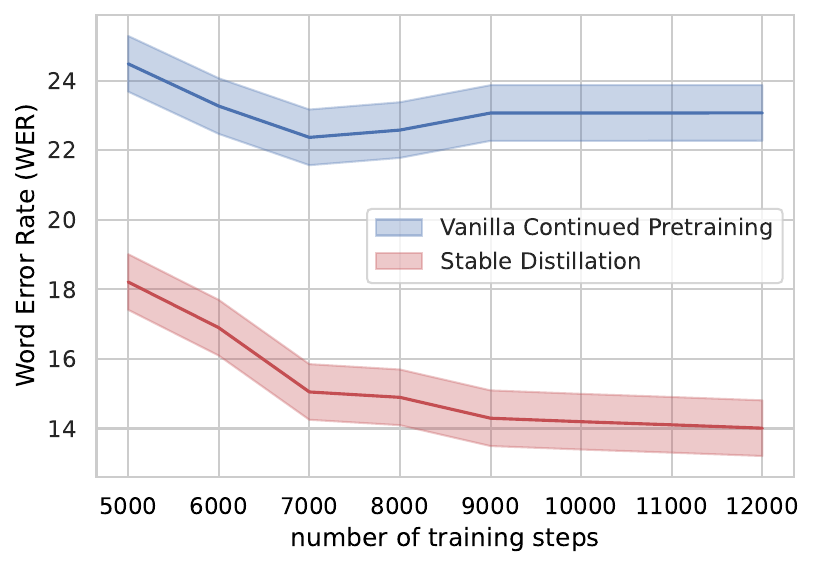}
    \caption{\small SwitchBoard test-set WER evaluated at various steps
    of continued SSL pre-training of a model already pre-trained with LibriSpeech. While vanilla CP starts over-fitting with increasing steps, Stable Distillation avoids it by regularization.}
    \label{fig:graph}
\end{figure}

Self-supervised learning (SSL) for learning effective speech representations without expensive labeled data has proven to be extremely successful for a variety of downstream Spoken Language Processing (SLP) tasks \cite{yang2021superb}. The general formulation of the training pipeline is to first perform large-scale upstream pre-training of a transformer model by solving a pretext task using unlabeled speech data and then fine-tune them on labeled datasets for diverse downstream tasks. The pre-text task to be solved can vary, and researchers have continuously found new state-of-the-art methods that push performance on low-resource SLP tasks~\cite{baevski2020wav2vec,hsu2021hubert} Despite its huge success in low-resource ASR, prior research has shown that SSL models suffer from various drawbacks: (1) Speech representation models learned with SSL often don't generalize well to domain shifts between pre-training and fine-tuning data \cite{hsu2021robust}, and (2) directly fine-tuning SSL models on downstream data often leads to catastrophic forgetting if there is a significant discrepancy between the pre-training and fine-tuning domains \cite{chen2020recall,aghajanyan2020better}. To overcome this, researchers have proposed \textit{continued SSL pre-training} (CP) on the downstream data to be an effective and complementary solution to bridge upstream pre-training and downstream fine-tuning domains \cite{gururangan2020don}, especially when the target domain dataset is low-resource. However, CP of an SSL model often leads to over-fitting on the train set~\cite{purushwalkam2022challenges}. Fig. \ref{fig:graph} shows how, with increasing steps of continued pre-training on a target ASR dataset, the test Word Error Rate (WER) keeps increasing. Over-fitting patterns in the target domain also lead the model to forget useful general-purpose knowledge learned in the past, leading to sub-optimal performance. While it is well known in the deep learning community that training on low-resource data can almost always lead a model to overfit on the train set and poorly generalize, Purushwalkam \textit{et al.}~\cite{purushwalkam2022challenges} discuss in detail why CP leads to over-fitting and conclude that the key reason is the violation of the IID assumption of optimization algorithms \cite{bousquet2003introduction}.


\begin{figure*}[t]
    \centering
    \includegraphics[width=0.9\textwidth]{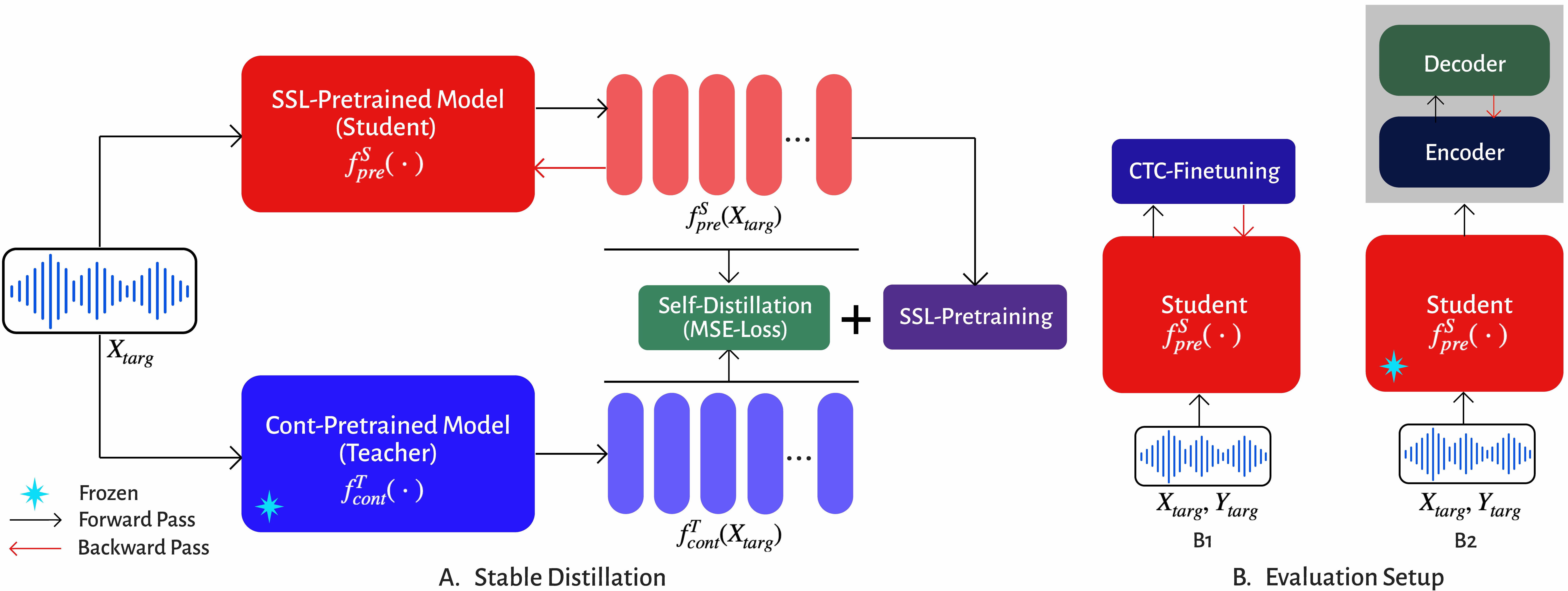}
    \caption{\small Illustration of Stable Distillation (SD). SD performs self-distillation with a teacher model while pre-training a student on the target ASR dataset. Both the student and the teacher models start from identical SSL pre-trained models, pre-trained on a large-scale OOD dataset. The teacher is additionally pre-trained on the target dataset before SD. For evaluation, the student model is either fine-tuned end-to-end using CTC (B$_1$) or used as a feature extractor for fine-tuning an Encoder-Decoder model (B$_2$).}
    \label{fig:sd_pic}
\end{figure*}

{\noindent \textbf{Main Contributions.}} In this paper, we present Stable Distillation, a simple and novel continued pre-training strategy to improve ASR performance on the target domain. Our key goal is to learn representations that, while adapting to the target domain, have not over-fitted the training dataset and are not forgetful of knowledge acquired during the initial pre-training. Stable Distillation, similar to other model distillation methods, employs two individual models, a teacher $f_{pre}^T(.)$ and a student $f_{pre}^S(.)$, both of which, in our case, are identical and pre-trained with SSL on a large-scale out-of-domain dataset. We first train the teacher with vanilla continued pre-training on the target downstream ASR data and solve the same pre-text task as the original SSL model. Next, we use the teacher to regularize continued pre-training on the student through self-distillation. Specifically, we attempt to match the representations of the student and the teacher by optimizing the MSE loss between both representations. Intuitively, the additional self-distillation regularizes the distance between the initial weight before continued pre-training and the final weight after fine-tuning. Finally, we either fine-tune the student for ASR or use it as a feature extractor. We conduct extensive experiments on several downstream datasets where the domain differs from the actual pre-training data domain and show that Stable Distillation can achieve absolute WER improvements in the range of 0.8 - 7 WER over our baselines that went through no or vanilla continued pre-training. Additionally, we show an implicit benefit of Stable Distillation where our method can be used for better adapting a cross-lingual model for mono-lingual ASR fine-tuning.




\section{Related Work}
{\noindent \textbf{SSL in Speech.}} The last couple of years have seen a massive rise in SSL research with models constantly outperforming SOTA on benchmark datasets \cite{panayotov2015librispeech,yang2021superb}. The most common pre-text tasks include instance discrimination via contrastive learning \cite{baevski2020wav2vec}, clustering \cite{hsu2021hubert}, or reconstruction \cite{liu2020mockingjay}. However, despite its success, SSL models are not robust to domain shifts between pre-training and fine-tuning datasets \cite{hsu2021robust}. This calls for better and more effective ways to adapt a pre-trained model to a target domain for downstream fine-tuning. Continued pre-training on the downstream dataset has shown to be an effective solution for target domain adaptation. However, it comes at a risk of over-fitting \cite{purushwalkam2022challenges}. Overall, through widely employed, continued pre-training of SSL models is a relatively under-explored area of speech representation learning, and we argue that more attention is required by the research community to solve problems related to it.

{\noindent \textbf{Domain Adaptation.}} Supervised domain adaption for ASR is a well-studied problem in literature \cite{meng2019domain}. On the contrary, unsupervised SSL model domain adaptation to the target task dataset domain is a relatively understudied problem in literature. Fan \textit{et al.} \cite{fan2022draft} were the first to explore this task and proposed DRAFT, which uses adapters in pre-trained SSL models for effective continued pre-training. Though adapters help effectively mitigate the problem of overfitting and catastrophic forgetting \cite{he2021effectiveness}, the need and complexity of adding extra parameters to the model impede their widespread usage. Lodagla \textit{et al.} \cite{lodagala2023pada} propose a pruning-assisted domain adaptation of SSL representations. However, their framework is complex, requires multiple pre-training and fine-tuning stages, and achieves minimal improvement in performance. 
\vspace{1mm}




\section{Methodology}
\label{sec:methodology}



{\noindent \textbf{Problem Formulation.}} Fig. \ref{fig:sd_pic} illustrates our proposed Stable Distillation. Our problem statement is simple. Given an upstream pre-trained model $f_{pre}(.)$, pre-trained on any unlabeled dataset $\mathcal{D}_p$ = ($\mathbf{X}_{pre}$) using SSL, we would like to employ $f_{pre}(.)$ to learn ASR on a downstream labeled dataset $\mathcal{D}_f$= ($\mathbf{X}_{targ}$,$\mathbf{Y}_{targ}$). For this, we either fine-tune the upstream SSL model using CTC or use it as a feature extractor to fine-tune a conformer-based Encoder-Decoder model to finally obtain an ASR model $\mathcal{F}$. As discussed earlier, though continued pre-training of $\mathcal{D}_f$ on $f_{pre}(.)$ can lead to optimal performance due to the several advantages it offers, like domain adaption to the target domain, but it also suffers from various drawbacks like over-fitting. Thus, we seek an effective solution for CP via our proposed Stable Distillation.

\subsection{Stable Distillation}
\label{sec:stable_distillation}
 
{\noindent \textbf{(1) Continue Pre-training the Teacher.}} The first step towards performing Stable Distillation is to teach the teacher. We initialize our teacher by making a copy of our pre-trained model $f_{pre}(.)$ and denote it as $f_{pre}^T(.)$. $f_{pre}^T(.)$ is then continually pre-trained on $\mathcal{D}_p$, by solving the same pre-text task that $f_{pre}(.)$ was pre-trained on, to finally obtain $f_{cont}^T(.)$
\vspace{1mm}

{\noindent \textbf{(2) Self-Distillation as Regularization.}} As a second step, we initialize our student by making a copy of our pre-trained model $f_{pre}(.)$, and denote it as $f_{pre}^S(.)$. Next, parallel to performing continued pre-training on the student, we perform self-distillation between $f_{pre}^S(.)$ and $f_{cont}^T(.)$ by matching their final layer representations of downstream dataset $\mathcal{D}_f$ to finally obtain $f_{dist}^S(.)$. Thus, for stable distillation, we jointly optimize the MSE between student and teacher representations and the student's own pre-text SSL task loss as:
\begin{multline}
\label{eqn:info}
\mathcal{L}_{sd}=\lVert f_{pre}^S(\mathbf{X}_{targ})-f_{cont}^T(\mathbf{X}_{targ}))\rVert_{2}\\
    +\alpha \mathcal{L}_{pretext}(f_{pre}^S(\mathbf{X}_{targ}))
\end{multline}
where $\alpha$ is a tunable hyper-parameter used to weigh the pre-text loss $\mathcal{L}_{pretext}$. The pre-text task is similar to the original.

\vspace{1mm}

{\noindent \textbf{(3) Final downstream fine-tuning.}} Finally, we take the trained student model $f_{dist}^S(.)$ and fine-tune it on the target dataset $\mathcal{D}_f$ using either: (1) End-to-End CTC Fine-tuning: In a more conventional setup, we fine-tune all the weights of $f_{dist}^S(.)$ by adding a linear CTC head and optimizing the CTC loss. (2) We fine-tune a conformer-based encoder-decoder model employing $f_{dist}^S(.)$  as an upstream feature extractor by jointly optimizing CTC and attention-based auto-regressive loss \cite{watanabe2017hybrid}. For the latter, all weights of $f_{dist}^S(.)$ are frozen.

\begin{figure}[t]
  \centering
  \includegraphics[width=0.35\textwidth]{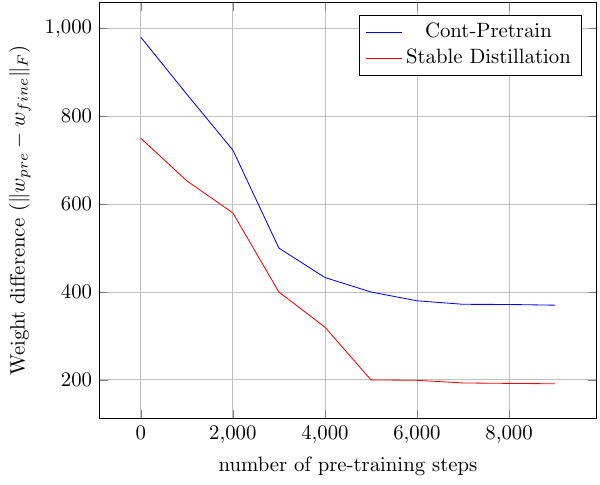}
  \caption{\small Frobenius norm of the weight difference between CP ($w_{pre}$) and fine-tuned models ($w_{fine}$) for CP of Wav2Vec2-Lb on SWBD.}
  \label{fig:graph_2}
\end{figure}

\vspace{1mm}
{\noindent \textbf{An intuition into why Stable Distillation works.}} Nagarjan \textit{et al.} \cite{nagarajan2019generalization} show that the model capacity of deep networks is restricted through implicit regularization of the $l_2$ distance from the initialization. Thus, the distance can affect the generalization bounds of these networks.
Mobahi \textit{et al.} \cite{mobahi2020self} perform several experiments to study the regularization effect of self-distillation. Thus, as a derivation of these findings, we use self-distillation to improve generalization via a regularization effect on the $l_2$ distance. Fig. \ref{fig:graph_2} illustrates how SD minimizes the $l_2$ distance between fine-tuned and CP models.

\begin{table}[t]
    \caption{\small Detailed Statistics of datasets used in our experiments. \textbf{Type} refers to Conversational or Read speech.}
    \centering
    \resizebox{1.0\columnwidth}{!}
    {\begin{tabular}{lcccc}
    \toprule
\textbf{Dataset}&\textbf{Language}&\textbf{Domain}&\textbf{Type}&\textbf{Duration}\\
    & & & &(train, dev, test)\\
    \toprule
    MSR & Gujarati & General& Conv.& 40hr, 5hr, 5hr\\
    MSR & Tamil & General& Conv.&40hr, 5hr, 5hr\\
    MSR & Telugu & General& Conv.&40hr, 5hr, 5hr\\
    Gramvani (GV) & Hindi & Call Cent.& Conv.&100hr, 5hr, 3hr\\
    SwitchBoard (SWBD) & English & Call Cent.& Conv. &30hr, 5hr, N.A.\\
    Wall Street Journal (WSJ) & English & Finance & Read & 80hr, 1.1hr, 0.4hr\\\hline
    
    \toprule
    
    \end{tabular}}
    \label{tab:my_label_dataset}
\end{table}

\begin{table*}[t]
\small
\centering
  \caption{\small Comparison of Stable Distillation ASR results with our baselines on both Enc-Dec and E2E evaluation settings. All results are in the format of \textbf{dev / test}. $\mathcal{R}$ and $\mathcal{C}$ indicate Read and Conversational Speech. Domain Map refers to the source pre-training $\rightarrow$ CP domain.}
  \label{tab:results}
 \resizebox{1.0\textwidth}{!}{
  \begin{tabular}{c c c c c c c c c}
    \toprule
    \textbf{Pretrained} & \textbf{Downstream} & \textbf{Domain Map} &  \multicolumn{2}{c}{\textbf{No Cont. Pretrain}} & \multicolumn{2}{c}{\textbf{Vanilla Cont. Pretrain}} & \multicolumn{2}{c}{\textbf{Stable Distillation}}\\
    \textbf{Model}& \textbf{Dataset} & (\emph{Source} $\rightarrow$ \emph{Target}) & \cellcolor{orange!20} Enc-Dec & \cellcolor{cyan!20} E2E & \cellcolor{orange!20} Enc-Dec & \cellcolor{cyan!20} E2E & \cellcolor{orange!20} Enc-Dec & \cellcolor{cyan!20} E2E\\
    \toprule
    XLSR-300 & GV\textsubscript{Hindi} & $\text{General}_\mathcal{R} \rightarrow \text{Call Cent.}_\mathcal{C}$ & \cellcolor{orange!20} 32.7 / 32.5 & \cellcolor{cyan!20} 37.3 / 37.0 & \cellcolor{orange!20} 31.6 / 31.4 & \cellcolor{cyan!20} 35.3 / 35.0 & \cellcolor{orange!20}\textbf{28.9 / 27.7} & \cellcolor{cyan!20}\textbf{30.3 / 30.1}\\
    XLSR-300 & MSR\textsubscript{Gujarati} & $\text{General}_\mathcal{R} \rightarrow \text{General}_\mathcal{C}$ & \cellcolor{orange!20} 21.7 / 28.5 & \cellcolor{cyan!20} 24.4 / 32.3 & \cellcolor{orange!20} 21.3 / 27.2 & \cellcolor{cyan!20} 22.1 / 30.3 & \cellcolor{orange!20} \textbf{20.2 / 26.4} & \cellcolor{cyan!20} \textbf{20.2 / 28.4}\\
    XLSR-300 & MSR\textsubscript{Tamil} & $\text{General}_\mathcal{R} \rightarrow \text{General}_\mathcal{C}$ & \cellcolor{orange!20} 28.1 / 27.7 & \cellcolor{cyan!20} 33.4 / 32.1 & \cellcolor{orange!20} 27.8 / 26.9 & \cellcolor{cyan!20} 32.2 / 31.2 & \cellcolor{orange!20} \textbf{26.7 / 25.7} & \cellcolor{cyan!20} \textbf{29.1 / 28.4}\\
    XLSR-300 & MSR\textsubscript{Telugu} & $\text{General}_\mathcal{R} \rightarrow \text{General}_\mathcal{C}$ & \cellcolor{orange!20} 28.3 / 28.8 & \cellcolor{cyan!20} 34.1 / 32.8 & \cellcolor{orange!20} 28.0 / 28.3 & \cellcolor{cyan!20} 32.6 / 32.0 & \cellcolor{orange!20} \textbf{27.2 / 27.1} &\cellcolor{cyan!20} \textbf{29.4 / 28.1}\\
    Vakyansh & GV\textsubscript{Hindi} & $\text{General}_\mathcal{R} \rightarrow \text{Call Cent.}_\mathcal{C}$ & \cellcolor{orange!20} 34.5 / 34.3 & \cellcolor{cyan!20} 33.2 / 34.2 & \cellcolor{orange!20} 32.7 / 32.5 & \cellcolor{cyan!20} 31.7 / 31.5 & \cellcolor{orange!20} \textbf{30.6 / 30.4} & \cellcolor{cyan!20} \textbf{30.0 / 30.1}\\
    Wav2Vec2-Lb & SWBD\textsubscript{English} & $\text{General}_\mathcal{R} \rightarrow  \text{Call Cent.}_\mathcal{C}$ & \cellcolor{orange!20} 39.1 / N.A & \cellcolor{cyan!20} 22.2 / N.A. & \cellcolor{orange!20} 36.2 / N.A. & \cellcolor{cyan!20} 20.4 / N.A. & \cellcolor{orange!20} \textbf{31.4 / N.A.} & \cellcolor{cyan!20} \textbf{14.3 / N.A.}\\
    Wav2Vec2-Lb & WSJ\textsubscript{English} & $\text{General}_\mathcal{R} \rightarrow \text{Finance}_\mathcal{R}$ & \cellcolor{orange!20} 12.4 / 11.6 & \cellcolor{cyan!20} 11.4 / 10.9 & \cellcolor{orange!20} 11.3 / 10.6 & \cellcolor{cyan!20} 10.5 / 10.0 & \cellcolor{orange!20} \textbf{10.5 / 9.8} &\cellcolor{cyan!20} \textbf{9.5 / 9.1}\\
    \bottomrule    
        
  \end{tabular}
  }
\end{table*}

\section{Experimental Setup}
\label{sec:experimental}

{\noindent \textbf{Datasets and SSL Pre-trained models.}} Details on individual datasets used for all our experiments can also be found in Table \ref{tab:my_label_dataset}. We use the splits mentioned in Table \ref{tab:my_label_dataset} for all our experiments in Table \ref{tab:results}. Precisely, we evaluate our approach on the MSR low resource Indian language corpus (MSR) \cite{inproceedings}, Gramvani ASR dataset \cite{bhanushali2022gram}, SwitchBoard dataset~\cite{godfrey1992switchboard} (SWBD) and the Wall Street Journal dataset~\cite{paul1992design}. For SWBD, we take a low-resource split, the rationale behind which is mentioned in Section \ref{sec:results}.


For all our experiments in Table \ref{tab:results}, we use either of: (1) Wav2Vec2 \cite{baevski2020wav2vec} pre-trained on 960 hours of LibriSpeech \cite{panayotov2015librispeech} (Wav2Vec2-Lb), (2) XLSR-300 model \cite{conneau2020unsupervised} pre-trained on a combination of VoxPopuli, MLS, CommonVoice, BABEL, and VoxLingua107, or (3) Vakyansh model \cite{gupta2021clsril} pre-trained on speech from Hindi YouTube videos. All these models are pre-trained on \textit{read} speech from the \textit{general} domain.
\vspace{1mm}

{\noindent \textbf{Baselines.}} Due to the lack of prior work in this space, we compare Stable Distillation with other commonly used pipelines that include: (1) No Continued Pre-training: For this baseline, we use the pre-trained SSL model without any continued pre-training on downstream dataset $\mathcal{D}_f$ and 
(2) Vanilla Continued Pre-training: For this baseline, we perform vanilla continued pre-training of our SSL model on downstream dataset $\mathcal{D}_f$.
\vspace{1mm}

{\noindent \textbf{Hyper-parameters.}} For stable distillation, we first perform continued pre-training on our teacher wav2vec-2.0 for a total of 50 epochs and then perform Stable Distillation on the student for 50 more epochs. For a fair comparison, for our continued pre-training baseline, we pre-train the model for a total of 50 epochs. We train with a learning rate of $5e^{-4}$ using Adam optimizer. Finally, for Stable distillation, we find $\alpha$ = 0.01 to give us the best performance searched among $\{0.01, 0.05, 0.1, 0.5, 1\}$. For downstream fine-tuning, we use a conformer-based encoder-decoder model that has 12 encoder layers and 6 decoder layers. We train our models with a learning rate of $1.5e^{-3}$, batch size of 64, and for a total of 100 epochs. We report the Word Error Rate (WER) averaged over three runs with different random seeds. 


\section{Results and Analysis}
\label{sec:results}

\begin{table}[t]
\small
\caption{\small Performance comparison in Enc-Dec setting when CP $\rightarrow$ final fine-tuning language/domain differs. Results show that SD leads to better regularization. For Libri. we report clean $\vert$ other.}
    \centering
    \resizebox{1.0\columnwidth}{!}{\begin{tabular}{lcccc}
    \toprule
    \textbf{Model}&\textbf{Dataset}&\textbf{Vanilla Cont. Pretrain}&\textbf{Stable Distillation}\\
    \toprule
    XLSR-300 & GV\textsubscript{Hindi} $\rightarrow$ MSR\textsubscript{Tamil} & \cellcolor{orange!20} 101.4 / 101.5 & \cellcolor{orange!20} \textbf{47.8 / 47.8}\\
    XLSR-300 & GV\textsubscript{Hindi} $\rightarrow$ MSR\textsubscript{Telugu} & \cellcolor{orange!20} 99.4 / 99.7 & \cellcolor{orange!20} \textbf{46.7 / 46.9}\\
    Wav2Vec2-Lb & WSJ\textsubscript{English} $\rightarrow$ Libri\textsubscript{English} & \cellcolor{orange!20} 18.2 / 28.8 $\vert$ 23.3 / 29.8 & \cellcolor{orange!20} \textbf{16.0 / 25.2 $\vert$ 20.1 / 27.6}\\

    \toprule
    
    \end{tabular}}
    
    \label{tab:my_label_3}
\end{table}

Table \ref{tab:results} compares the performance of Stable Distillation with all our baselines on the splits for all datasets mentioned in Table \ref{tab:my_label_dataset}. For all experiments, we follow dataset settings resembling real-world scenarios where the target domain for CP is usually from a different domain and is low-resource compared to the source domain that the model is already pre-trained on (see Domain Map column in Table \ref{tab:results}). SD, on average, outperforms our baselines by 0.8-7.7 WER in the Enc-Dec setup and 1-7 WER in the E2E setup. Precisely, for the Enc-Dec setup, compared to vanilla CP, we achieved an average relative WER improvement of 7.2\% and 7.1\% for dev and test sets, respectively. Compared to no CP, the improvements are 11.3\% 11.1\%. For the E2E setup compared to vanilla CP, we achieved an average relative WER improvement of 13.0\% and 12.9\% on dev and test sets, respectively. Compared to no CP, the improvements are 17.3\% and 17.2\%.\\
Table \ref{tab:my_label_3} confirms our claim that SD improves vanilla CP by adding explicit regularization, thereby preventing overfitting on the target domain train set. Ideally, if a model has overfitted to the target training dataset after CP, it should perform worse if it is fine-tuned for ASR on the source dataset itself or any other dataset where the domain or language differs. While rows 1 and 2 show examples when the CP and fine-tuning languages differ, row 3 shows that SD leads to better performance than vanilla CP with LibriSpeech ASR when Wav2Vec2-Lb (also pre-trained on LibriSpeech) was further pre-trained on SWBD. Row 6 in Table \ref{tab:results} shows that the same model also performs better than CP with WSJ ASR (due to effectively exploiting past knowledge).  Rows 1-6 also show a peripheral benefit of SD, which helps in effective cross-lingual to mono-lingual adaptation of SSL models pre-trained on large-scale multi-lingual data~\cite{khurana2022magic}.

\begin{figure}[t] 
    \centering 
    
    \begin{subfigure}{0.23\textwidth} 
        \centering
        \includegraphics[width=\linewidth]{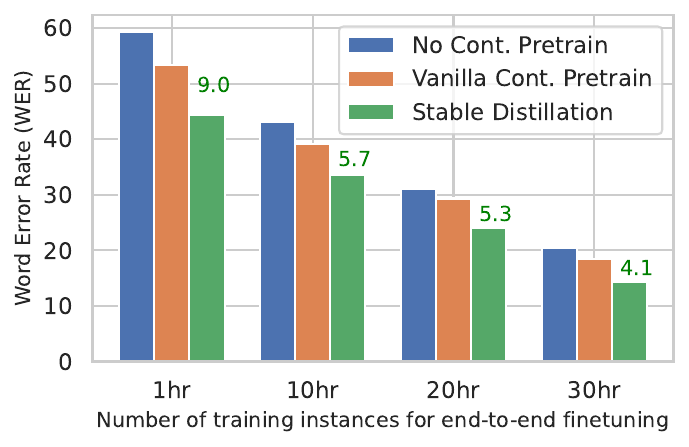}
        \caption{\small Evaluation-set WER on SWBD on various low-resource settings for E2E setup.}
        \label{fig:subfig1}
    \end{subfigure}
    \hfill 
    \begin{subfigure}{0.23\textwidth}
        \centering
        \includegraphics[width=\linewidth]{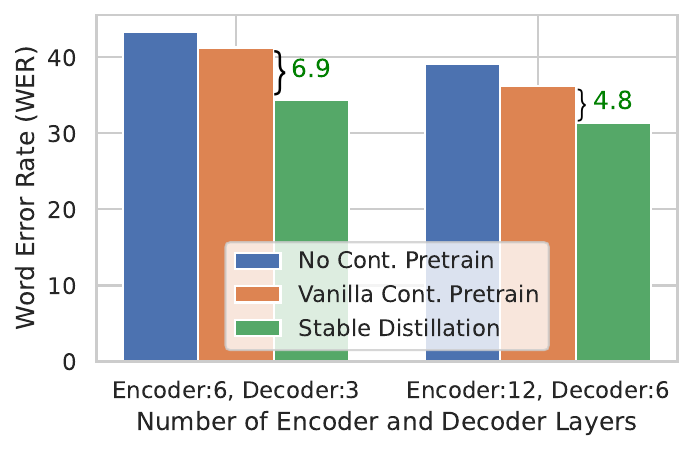}
        \caption{\small Evaluation-set WER on SWBD for various Enc-Dec layer configs on Wav2Vec2-Lb.}
        \label{fig:subfig2}
    \end{subfigure}
    
    \caption{\small More results. Stable Distillation outperforms all baselines.}
    \label{fig:mainfig}
\end{figure}

\section{Conclusion and Future Work}
In this paper, we propose Stable Distillation, a simple and effective methodology for target domain adaptation of pre-trained SSL models. Stable Distillation regularizes continued pre-training of an SSL model on a target domain and prevents overfitting. We empirically show that Stable Distillation proves to be extremely effective in various settings where the initial and target domains differ. As part of future work, we would like to explore better distillation procedures for continued pre-training further to push ASR performance in trivial and simple steps.


\bibliographystyle{IEEEbib}
\bibliography{strings,refs}

\end{document}